\documentclass[prl,amsmath,twocolumn]{revtex4-2}
\usepackage[utf8]{inputenc}

\usepackage{amssymb}
\usepackage{amsmath}
\usepackage{amsfonts}
\usepackage{bbold}
\usepackage{bm}
\usepackage{bbm}
\usepackage{braket}
\usepackage{comment}
\usepackage{color}
\usepackage{epsfig}
\usepackage{graphicx}
\usepackage{hyperref}
\usepackage{listings}
\usepackage{soul}
\usepackage{tikz}
\usepackage{multirow}
\usepackage[linesnumbered,ruled]{algorithm2e}

\newcommand{\nn}{\nonumber}
\newcommand{\ud}{{\textrm{d}}}

\begin{document}

\title{Disorder-induced spin-charge separation in the 1-D Hubbard model}
\author{S.~J.~Thomson}
\affiliation{Dahlem Centre for Complex Quantum Systems, Freie Universität, 14195 Berlin, Germany}
\date{\today}

\begin{abstract}
    Many-body localisation is believed to be generically unstable in quantum systems with continuous non-Abelian symmetries, even in the presence of strong disorder. Breaking these symmetries can stabilise the localised phase, leading to the emergence of an extensive number of quasi-locally conserved quantities known as local integrals of motion, or $l$-bits. Using a sophisticated non-perturbative technique based on continuous unitary transforms, we investigate the one-dimensional Hubbard model subject to both spin and charge disorder, compute the associated $l$-bits and demonstrate that the disorder gives rise to a novel form of spin-charge separation. We examine the role of symmetries in delocalising the spin and charge degrees of freedom, and show that while symmetries generally lead to delocalisation through multi-particle resonant processes, certain subsets of states appear stable.
\end{abstract}

\maketitle

\emph{Introduction} - 
The study of localisation in low-dimensional quantum systems due to disorder has a long history, from the original works of Anderson~\cite{Anderson58,Fleishman+80} through to modern ideas of many-body localisation~\cite{Basko+06,Pal+10,Huse+13,Luitz+15,Nandkishore+15,Altman+15,Alet+18,AbaninEtAlRMP19}.
Conventionally, many-body localisation (MBL) in systems with a random disordered potential is now understood in terms of the existence of an extensive number of local integrals of motions (LIOMs, or $l$-bits) which decay exponentially in space and are related to the constituents of the microscopic model via a quasi-local unitary transform~\cite{SerbynPapicAbaninPRL13_2,HuseNandkishoreOganesyanPRB14,Imbrie16a,Ros+15,Imbrie+17}. Several methods for constructing these $l$-bits exist, including computing them directly via unitary transforms~\cite{Rademaker+16,Monthus16,Rademaker+17,Pekker+17,Thomson+18}, extracting them from the long-time behaviour of various observables~\cite{Chandran+15,Goihl+18} and from other hydrodynamic considerations~\cite{Singh+21}.

Much of the focus has been on spinless fermionic systems or spin-1/2 chains, however experiments~\cite{Kondov+15,Schreiber+15,Bordia+17} have also studied localisation in systems of \emph{spinful fermions} which exhibit an $\mathrm{SU}(2)$ spin rotation symmetry. It has been argued that full many-body localisation cannot exist unless all continuous non-Abelian symmetries are broken~\cite{Mondaini+15,Potter+16,Prelovsek+16,Protopopov+17,Kozarzewski+18,Mierzejewski+18,Zakrewski+18,Leipner-Johns+19,Sroda+19,Protopopov+19,Suthar+20}, and that even then, not all eigenstates may display typical MBL phenomena~\cite{Yu+18}. Models of spinful fermions present a major challenge to exact numerical methods due to the additional degree of freedom compared with spinless fermions, and this places strong limitations on the system sizes which can be studied numerically. 
In particular, little is known about the form of $l$-bits in systems of spinful fermions, despite their experimental relevance and the central role that $l$-bits play in describing and understanding many-body localised quantum matter.
The study of MBL in spinful fermionic systems is also relevant for the broader question of the stability of MBL, as the interaction between fermionic species can be viewed as a pseudo `system-bath' coupling.

In this work, we make use of a technique designed to compute the local integrals of motion associated to a model with multiple fermionic species, and show that the existence of any continuous non-Abelian symmetry prevents the Hamiltonian from being written as a sum of mutually commuting local integrals of motion. When the symmetry is broken, we demonstrate that it is possible to write an effective Hamiltonian describing the system in terms of local integrals of motion associated to charge and spin degrees of freedom. This is reminiscent of the spin-charge separation seen in various regimes of the disorder-free Hubbard model -- notably in the Luttinger liquid regime~\cite{Haldane81,Kollath+05,Jompol+09}, the related strongly interacting $t$-$J$ model~\cite{Spalek88,Wang+94,Putikka+94,Chen+94,Weng+95,Vijayan+20} as well as in novel driven phases~\cite{Gao+20} -- but here it arises solely due to the presence of disorder~\cite{Zakrewski+18}. 

\begin{figure}[t!]
    \centering
    \includegraphics[width=0.9\linewidth]{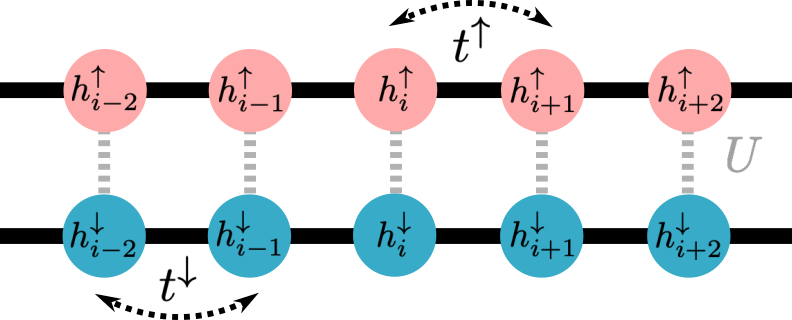}
    \caption{A sketch of the Hubbard model described in Eq.~\ref{eq.Hubbard} as two coupled chains of spinless fermions subject to a disordered on site potential $h_i^{\sigma}$ and with hopping amplitude $t^{\sigma}$, interacting via an on-site repulsion $U$.
    \vspace{-3mm}}
    \label{fig.cartoon}
\end{figure}

\emph{Model} - 
We shall study the behaviour of spinful fermions on a lattice in one dimension using the Fermi-Hubbard model, sketched in Fig.~\ref{fig.cartoon}, which is given by:
\begin{align}
\mathcal{H} &= \sum_i \sum_{\sigma=\uparrow,\downarrow} h_i^{\sigma} n_{i,\sigma} - \sum_{i} \sum_{\sigma=\uparrow,\downarrow} t^{\sigma} (c^{\dagger}_{i,\sigma} c_{i+1,\sigma} + H.c.) \nn \\
& \quad \quad \quad + U \sum_i n_{i,\uparrow} n_{i,\downarrow},
\label{eq.Hubbard}
\end{align}
where $h_i^{\sigma} \in [-d,d]$ represents the (spin-dependent) on-site disorder potential. Throughout, we shall use $t^{\sigma} \equiv t=1.0$ as the unit of energy, and set $U/t=0.1$ unless otherwise stated. The disorder-free ($d=0$) model has an $\mathrm{SU}(2)$ symmetry associated to global spin rotations, as well as an $\mathrm{SU}(2)$ pseudospin symmetry associated with charge degrees of freedom~\cite{Yang89,Zhang90,Shen+96,Yu+18,Boretsky+18,Moudgalya+20}. We shall consider two types of disorder. The case of $h_i^{\uparrow}=h_i^{\downarrow}$ represents disorder in the charge degrees of freedom only, which preserves the $\mathrm{SU}(2)$ spin symmetry but breaks the $\mathrm{SU}(2)$ pseudospin symmetry, while the opposite case of $h_i^{\uparrow}=-h_i^{\downarrow}$ represents spin disorder which breaks the $\mathrm{SU}(2)$ spin symmetry but preserves the pseudospin symmetry. 
In the non-interacting case of $U/t=0$, the spin-up and spin-down fermions become completely decoupled and one can imagine the Hamiltonian Eq.~\ref{eq.Hubbard} as describing two independent chains of non-interacting particles. In the presence of disorder ($h_i^{\uparrow} \neq 0,h_i^{\downarrow} \neq 0$), these non-interacting chains will both be Anderson localized. 

\emph{Method} - 
We diagonalize the Hamiltonian up to terms of quartic order in the fermionic operators using the Tensor Flow Equation (TFE) method~\cite{Thomson+21,SM}, implemented using \texttt{PyFlow}~\cite{PyFlow} and building on a variety of earlier works~\cite{Thomson+18,Thomson+20,Thomson+20b,Thomson+20c}. It is based on the G\l{}azek-Wegner-Wilson flow, a method that has been widely used in condensed matter physics (and beyond) for a large variety of problems~\cite{Brockett91,Chu94,Wegner94,Glazek+93,Glazek+94,Stein97,Mielke98,Knetter+00,Knetter+03,Kehrein07,Yang+12,Monthus16,Quito+16,Pekker+17,Kelly+20}, including time-dependent~\cite{Tomaras+11,Verdeny+13,Vogl+19,Thomson+20c} and open~\cite{Rosso+20} systems. This technique is non-perturbative~\cite{Kehrein07}, but its success in fully diagonalising the Hamiltonian depends on the microscopic model having a well-defined separation of energy scales. The method works by successively applying infinitesimal unitary transforms to the Hamiltonian until convergence is achieved, at each step removing some fraction of off-diagonal terms at the cost of inducing new higher-order terms proportional to powers of the interaction strength. 
In order to keep these terms under control, we shall mainly work at weak interactions $U/t \ll 1$, such that any newly generated terms are small and can be neglected. A full error analysis is available in~\cite{SM}. Here, we will focus on the qualitative form of the transformed Hamiltonian:
\begin{align}
    \tilde{\mathcal{H}} &= \sum_i (\tilde{h}_i^{\uparrow} :\tilde{n}_{i,\uparrow}: + \tilde{h}_i^{\downarrow} :\tilde{n}_{i,\downarrow}:) + \sum_{ij} + \tilde{\Delta}_{ij} :\tilde{n}_{i,\uparrow} \tilde{n}_{j,\downarrow}:  \nn \\
    &\quad + \sum_{ij} \left( U^{*}_{ij} :\tilde{c}_{i,\uparrow}^{\dagger} \tilde{c}_{j,\uparrow} \tilde{c}_{i,\downarrow}^{\dagger} \tilde{c}_{j,\downarrow}:   + \overline{U}_{ij} :\tilde{c}_{i,\uparrow}^{\dagger} \tilde{c}_{j,\uparrow} \tilde{c}_{j,\downarrow}^{\dagger} \tilde{c}_{i,\downarrow}: \right)\nn \\
    &\quad + \sum_{ij} \left( \tilde{\Delta}_{ij}^{\uparrow} :\tilde{n}_{i,\uparrow} \tilde{n}_{j,\uparrow}: + \tilde{\Delta}_{ij}^{\downarrow} :\tilde{n}_{i,\downarrow} \tilde{n}_{j,\downarrow}: \right)  + ...  \label{eq.H_FP}
\end{align}
where the $...$ refers to higher-order terms which are neglected within the approximations made here, the $:O:$ notation signifies normal-ordering, and the tilde notation indicates that all quantities are expressed in the transformed basis. The form of this Hamiltonian is independent of the dimensionality of the system.

A few comments are in order about the form of this effective Hamiltonian. Firstly, and in contrast with previous implementations of the TFE method, this Hamiltonian is \emph{not} always fully diagonalised. The presence of the spin $\mathrm{SU}(2)$ symmetry leads to a degeneracy between spin-up and spin-down sectors of the Hamiltonian, which in turn leads to a spin flip term $\overline{U}_{ij}$. Similarly, the presence of the pseudospin $\mathrm{SU}(2)$ symmetry leads to a pair hopping term $U^{*}_{ij}$. If either symmetry is broken, and the resulting degeneracy lifted, the corresponding coupling constant is zero. These resonant terms couple degenerate states, and therefore cannot be removed by the TFE method. They do not commute with the rest of the Hamiltonian, ruling out the possibility of writing Eq.~\ref{eq.Hubbard} solely in terms of mutually commuting quasilocal integrals of motion (as removing real-space degeneracies in this framework typically requires a non-local unitary transform). We shall nonetheless refer to this as the `diagonal basis', as all single-particle off-diagonal and multi-particle off-diagonal, off-resonant terms have been removed, and refer to number operators in this basis as `$l$-bits'.
Secondly, the action of the flow equation method leads to the emergence of new interaction terms $\tilde{\Delta}^{\sigma}_{ij} \propto t U^2$ which act within each spin species: these terms arise due to non-perturbative corrections computed with respect to an appropriate reference state (here taken to be an excited state of the non-interacting system) via a normal-ordering procedure~\cite{SM}, and are in some cases equivalent to one-loop renormalisation group contributions~\cite{Kehrein07,Wegner06}. The computational cost of each step of this method is $\mathcal{O}(L^6)$, restricting us to small system sizes. For normal-ordering with respect to the vacuum state, the corrections are zero and $\tilde{\Delta}^{\sigma}_{ij}=0 \phantom{.} \forall \phantom{.} \sigma,i,j$. For normal-ordering with respect to an arbitrary excited state -- more appropriate for the study of many-body localisation -- these corrections are non-zero. This suggests that the low-energy and high-energy sectors of Eq.~\ref{eq.Hubbard} may behave differently in the presence of disorder. Here, we choose the reference state to be a normalised eigenstate of the non-interacting system which takes the form $\ket{0, \uparrow \downarrow,0, \uparrow \downarrow ...}$ in the diagonal basis~\cite{SM}. We assume that superconductivity plays no role (as we consider a disordered system far from the ground state) and consequently no anomalous terms appear in Eq.~\ref{eq.H_FP}.

Using the definition of charge ($\rho_i = :\tilde{n}_{i,\uparrow}: + :\tilde{n}_{i,\downarrow}:$) and spin ($\sigma_i = :\tilde{n}_{i,\uparrow}: - :\tilde{n}_{i,\downarrow}:$), we can (partially) rewrite this Hamiltonian in terms of mutually commuting $l$-bits associated with spin and charge degrees of freedom:
\begin{align}
    \tilde{\mathcal{H}} &= \sum_i \left(h_i \rho_i + \overline{h}_i \sigma_i  \right)+ \sum_{ij} \left( \Delta_{ij}^{\rho} \rho_i \rho_j + \Delta_{ij}^{\sigma} \sigma_i \sigma_j \right)  \label{eq.Hdiag} \\
    &+ \sum_{ij} ( \overline{U}_{ij} S^{+}_i S^{-}_j + U^{*}_{ij} P_i^{+}P_j^{-}) + \sum_{ij} \Omega_{ij} (\rho_i \sigma_j + \sigma_i \rho_j) + ... \nn
\end{align}
with $h_{i} = \frac12 (\tilde{h}_i^{\uparrow}+\tilde{h}_i^{\downarrow})$, $\overline{h}_{i} = \frac12 (\tilde{h}_i^{\uparrow}-\tilde{h}_i^{\downarrow})$, $\Delta_{ij}^{\rho} = \frac14 (\tilde{\Delta}^{\uparrow}_{ij}+\tilde{\Delta}^{\downarrow}_{ij}+\tilde{\Delta}_{ij})$, $\Delta_{ij}^{\sigma} = \frac14 (\tilde{\Delta}^{\uparrow}_{ij}+\tilde{\Delta}^{\downarrow}_{ij}-\tilde{\Delta}_{ij})$, $\Omega_{ij} = \frac14 (\tilde{\Delta}^{\uparrow}_{ij}-\tilde{\Delta}^{\downarrow}_{ij})$ and we define the spin-flip term $S^{+}_i S^{-}_j = :\tilde{c}_{i,\uparrow}^{\dagger} \tilde{c}_{j,\uparrow} \tilde{c}_{j,\downarrow}^{\dagger} \tilde{c}_{i,\downarrow}:$ and the pair hopping term $P^{+}_i P^{-}_j = :\tilde{c}_{i,\uparrow}^{\dagger} \tilde{c}_{j,\uparrow} \tilde{c}_{i,\downarrow}^{\dagger} \tilde{c}_{j,\downarrow}:$~\footnote{In general $:n_i n_j: \neq :n_i: :n_j:$, however in this case all lower-order normal-ordering terms evaluate to zero.}. 
The first line of Eq.~\ref{eq.Hdiag} represents two independent systems of `charge' and `spin' $l$-bits respectively, while the second line contains a coupling between them. In the case of charge disorder where only the spin $\mathrm{SU}(2)$ symmetry is preserved, we have $\overline{h}_i = 0 \phantom{.} \forall i$ and $U^{*}_{ij}=\Omega_{ij} = 0  \phantom{.} \forall i,j$, implying that in the absence of any other coupling all degrees of freedom would be localised. 
The spin flip coefficient $\overline{U}_{ij}$ is not zero, however, and this term does not commute with the $\sigma_i$ operators. This means that even though the charge degrees of freedom may be localised, the spin degrees of freedom are not. Similarly, in the case of spin disorder, the presence of an $\mathrm{SU}(2)$ pseudospin symmetry leads to the emergence of the pair hopping term $U^{*}_{ij} P_i^{+}P_j^{-}$, which does not commute with the $\rho_i$ operators, resulting in localisation of the spin degrees of freedom but not the charges.
In this case, however, both spin and charge degrees of freedom may eventually be able to equilibrate via the non-zero coupling $\Omega_{ij}$, although in practice we find that this term is small and decays exponentially with distance, and as such delocalisation via this term is likely to become extremely slow at strong disorder strengths. 
The spin flip term $S^{+}_i S^{-}_j$, when rewritten in terms of the original operators in the $l=0$ basis via the operator expansion $:\tilde{c}_{i,\sigma}^{\dagger}: = \sum_{j,\sigma'} \tilde{\alpha}_{j,\sigma'}^{(i,\sigma)} :c_{j,\sigma'}^{\dagger}: + \sum_{jkq, \sigma',\sigma''} \tilde{\beta}_{jkq,\sigma' \sigma''}^{(i,\sigma)} :c_{j,\sigma'}^{\dagger} c_{k,\sigma''}^{\dagger} c_{q,\sigma''}: + ...$~\cite{SM}, contains a spin-mediated hopping term of precisely the type obtained in Ref.~\cite{Protopopov+19}, as well as additional higher-order terms which combine to constitute a (slow) relaxation channel. The pair hopping term $P_i^{+}P_j^{-}$ has a similar effect. As the microscopic interaction strength $U$ is increased, both processes will become increasingly relevant for the dynamics, favouring thermalisation.
Curiously, there are also exceptional states where these terms can have no effect. Consider for example applying the spin-flip term to a CDW state in the diagonal basis, e.g. $\ket{\tilde{\psi}} = \ket{0, \uparrow \downarrow, 0, \uparrow \downarrow...}$ This term can only exchange pairs of opposite spins on different sites, and cannot modify the CDW state as double-occupancy of the same spin species is forbidden by fermionic statistics. Similarly, if we consider applying the pair hopping term to a spin density wave (SDW) state in the diagonal basis, $\ket{\tilde{\psi}} = \ket{\uparrow, \downarrow, \uparrow, \downarrow...}$, we find that it cannot change the state, as it can only move pairs of spins from one site to another. This suggests that for both types of disorder, although typical states will be delocalised, there may be rare states in which localisation is stable to long times, although we cannot rule out the existence of weak higher-order processes involving three or more particles which may destabilise these states on long timescales. This has clear parallels with weak ergodicity breaking, e.g. quantum many-body scars~\cite{Turner+18,Choi+19,Serbyn+21}, and is consistent with similar observations in Ref.~\cite{Yu+18}.
\begin{figure}[t!]
    \centering
    \includegraphics[width=\linewidth]{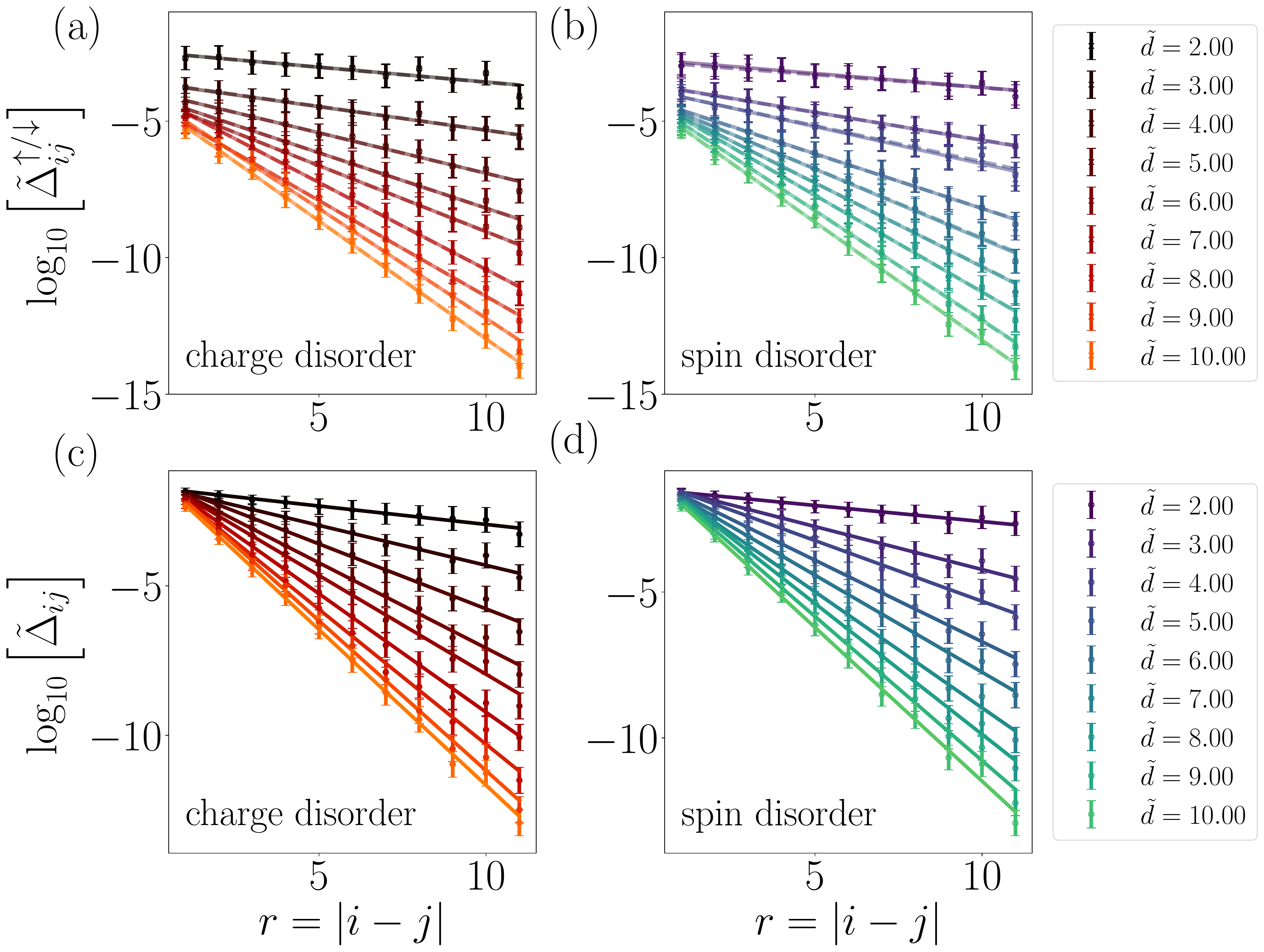}
    \caption{Coupling constants extracted from the diagonal Hamiltonian (Eq.~\ref{eq.H_FP}) for a system of size $L=12$ with $N_s=128$ in the case of charge disorder (left column) and spin disorder (right column), with $\tilde{d}=d/t$. a) Spin-up (crosses joined with solid lines) and spin-down (open circles joined with dashed lines) interactions. The solid and dashed lines are in most cases directly superimposed. b) The same, but here for spin disorder. c) The coefficient of the mixed spin interaction term $\tilde{\Delta}_{ij}$ in the case of charge disorder. d) The same again, but here for spin disorder. The decay of these interactions is essentially independent of the type of disorder used.
    \vspace{-4mm}}
    \label{fig1}
\end{figure}

\begin{figure}[t!]
    \centering
    \includegraphics[width=\linewidth]{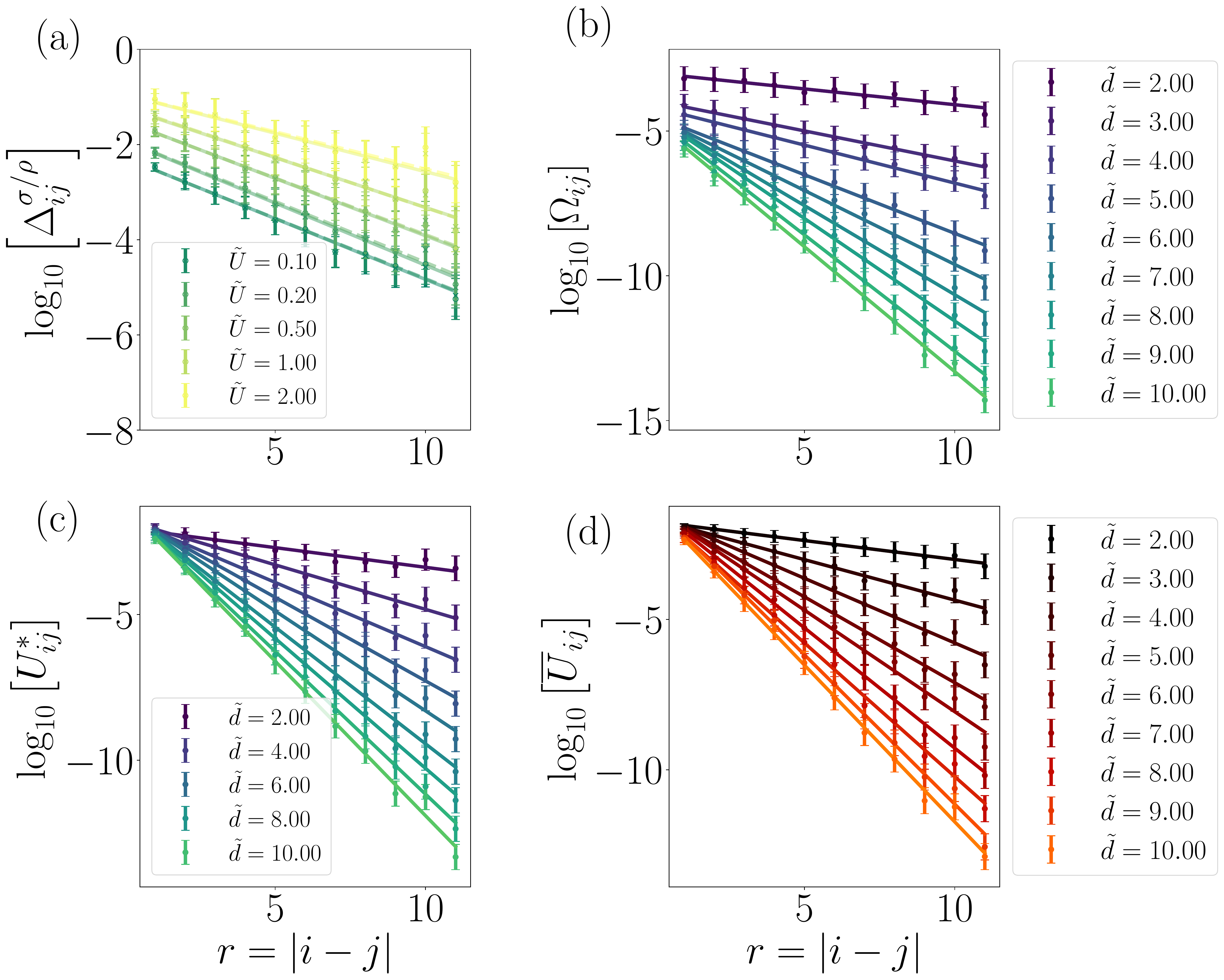}
    \caption{a) The coupling constants $\Delta_{ij}^{\rho}$ (solid lines) and $\Delta_{ij}^{\sigma}$ (dashed lines) extracted from Eq.~\ref{eq.Hdiag} at fixed (charge) disorder $d/t=3.0$ and varying interactions $\tilde{U}=U/t$.  b) The coefficient of the spin-charge coupling term in Eq.~\ref{eq.Hdiag}, for spin disorder and fixed interaction strength $U/t=0.1$ and varying disorder strengths $\tilde{d}=d/t$. c) The coefficient of the pair hopping term $U^{*}_{ij}$ in Eq.~\ref{eq.Hdiag}, again for spin disorder and fixed interaction strength while varying the disorder. d) The coefficient of the spin-flip term $\overline{U}_{ij}$ in Eq.~\ref{eq.Hdiag}, for charge disorder.
    \vspace{-5.5mm}}
    \label{fig2}
\end{figure}

\emph{Numerical Results} - 
The results of the flow equation procedure are shown in Fig.~\ref{fig1} for both charge and spin disorder. We will use a system of size $L=12$ and show the typical (median) magnitudes of various quantities $[ \hat{O} ]$ computed from $N_s \in[100,128]$ disorder realisations. Error bars represent the statistical uncertainty (median absolute deviation) unless otherwise stated, and all joining lines are guides to the eye. A few general features are worth commenting on. Firstly, the same-species interactions for spin-up (solid lines) and spin-down (dashed lines) $l$-bits behave almost indistinguishably regardless of the type of disorder used, in all cases exhibiting an exponential decay with distance. The mixed spin interaction term $\tilde{\Delta}_{ij}$ behaves similarly. Interestingly, these coupling constants are all largely independent of the presence of either $\mathrm{SU}(2)$ symmetry, meaning those effects must be solely felt elsewhere. Secondly, for small values of $d/t$, all coupling constants become highly extended, at which point the neglected high-order terms in Eq.~\ref{eq.Hdiag} become relevant and must be included. 
It is also interesting to see how the observed features change as the microscopic interaction strength $U$ is increased. In Fig.~\ref{fig2}a), we show the coupling constants $\Delta_{ij}^{\rho}$ and $\Delta_{ij}^{\sigma}$ for a fixed (charge) disorder strength of $d/t=3.0$ and for varying interaction strengths (with $N_s=100$). Both interactions increase with $U$. For $U \gtrsim 1$, the neglected higher-order terms in Eq.~\ref{eq.H_FP} will become relevant and the current approximation will break down, but the qualitative behaviour is nonetheless instructive and remains well-controlled.

To unveil the role of the $\mathrm{SU}(2)$ symmetries, we now turn to the coupling between spin and charge $l$-bits, namely the spin-charge coupling $\Omega_{ij}$,the pair hopping term $U^{*}_{ij}$, and the spin-flip term $\overline{U}_{ij}$ in Eq.\ref{eq.Hdiag}, shown in Fig.~\ref{fig2}b-d) respectively ($N_s=128$). The spin-flip term is only non-zero when the the $\mathrm{SU}(2)$ spin symmetry is preserved, while the spin-charge coupling term is only non-zero when this symmetry is broken, and the pair-hopping term is only non-zero in the presence of the $\mathrm{SU}(2)$ pseudospin symmetry.
Fig.~\ref{fig2}c) and d) show that for weak disorder the couplings $U^{*}_{ij}$ and $\tilde{U}_{ij}$ decay slowly and long-range resonances are possible, while at stronger disorder they are exponentially suppressed, suggesting delocalisation will become parametrically slow for strong disorder.

\begin{figure}[t!]
    \centering
    \includegraphics[width=\linewidth]{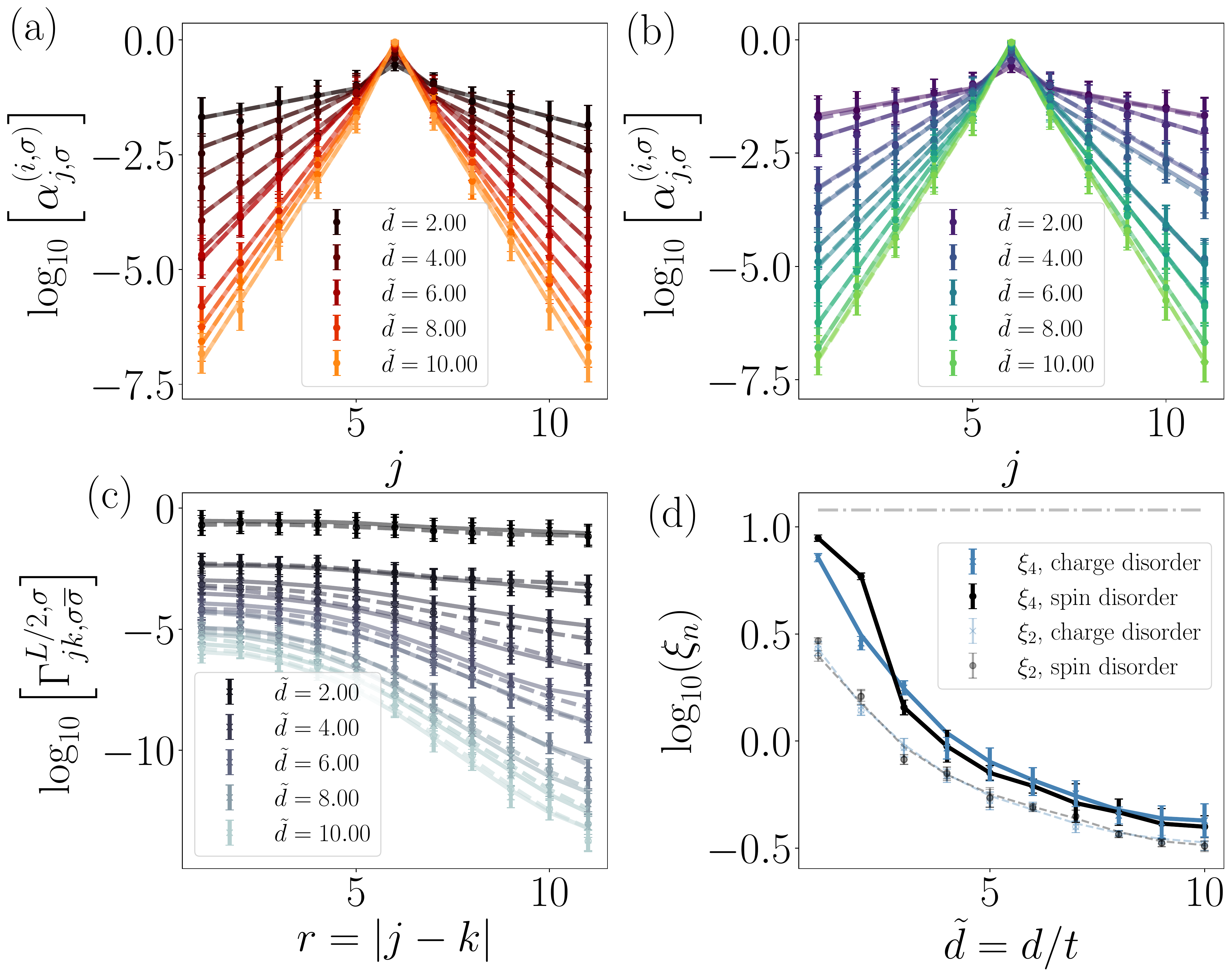}
    \caption{a) The quadratic terms $\alpha^{(i,\sigma)}_{j,\sigma}$ of Eq.~\ref{eq.LIOM} in the case of charge disorder, for $\sigma=\uparrow$ (solid lines) and $\sigma=\downarrow$ (dashed lines). b) The same quantity in the case of spin disorder. c) The quartic terms $\Gamma^{(i,\uparrow)}_{jk,\uparrow \downarrow} \equiv \zeta^{(i,\uparrow)}_{jjkk,\uparrow \downarrow}$ of Eq.~\ref{eq.LIOM} for charge disorder (solid) and spin disorder (dashed). d) Localisation lengths $\xi_2$ (dashed) and $\xi_4$ (solid), for charge disorder (blue) and spin disorder (black). Error bars indicate the fitting uncertainty. The grey dashed line indicates the system size, $L=12$.
    \vspace{-6mm}}
    \label{fig3}
\end{figure}

We can also directly compute the `integrals of motion' associated with the spin-up and spin-down fermions, using the TFE method to transform the local operators in Eq.~\ref{eq.H_FP} back into the original basis:
\begin{align}
    &:\tilde{n}_{i,\sigma}: = \sum_{j,\sigma'=\uparrow,\downarrow} \alpha^{(i,\sigma)}_{j,\sigma'} :n_{j,\sigma'}: + \sum_{jk,\sigma'} \beta^{(i,\sigma)}_{jk,\sigma'} :c^{\dagger}_{j,\sigma'} c_{k,\sigma'}: \nn \\
    &\quad + \sum_{jkpq,\sigma' \sigma''} \zeta^{(i,\sigma)}_{jkpq,\sigma' \sigma''} :c^{\dagger}_{j,\sigma'} c_{k,\sigma'} c^{\dagger}_{p,\sigma''} c_{q,\sigma''}: + ...
\label{eq.LIOM}
\end{align}
The quadratic terms $\alpha^{(i,\sigma)}_{j,\sigma}$ for both $\sigma=\uparrow,\downarrow$ are shown in Fig.~\ref{fig3}a-b) as a function of disorder strength, for both spin and charge disorder ($N_s=128$). They decay exponentially with distance $r=|i-j|$ and can be fitted with an exponential decay of the form $\sim \textrm{e}^{-r/\xi_2}$ to extract a localisation length $\xi_2$. Similarly, we can plot the quartic coefficient $\zeta^{(i,\sigma)}_{jkpq,\sigma' \sigma''}$ (the $\Gamma^{(i,\uparrow)}_{jk,\uparrow \downarrow} \equiv \zeta^{(i,\uparrow)}_{jjkk,\uparrow \downarrow}$ component is shown in  Fig.~\ref{fig3}c) - other components behave similarly), which also decays approximately exponentially at large distance for all combinations of distances (e.g. $|i-j|$, $|i-k|$, $|j-k|$ and so on) for any choice of $\sigma'$ and $\sigma''$, and we can fit the tails of these coefficients with a function $\sim \textrm{e}^{-r/\xi_4}$ to extract a second localisation length $\xi_4$. Both localisation lengths are shown in Fig.~\ref{fig3}d), and both exhibit similar monotonic decay with increasing disorder strength. At weak disorder, the localisation length $\xi_4$ associated to the interacting part of the $l$-bits becomes much larger then the non-interacting localisation length $\xi_2$ and approaches the system size, suggestive of interaction-driven delocalisation at sufficiently low values of $d/U$.

\emph{Discussion} - Starting from two decoupled, Anderson-localised chains of non-interacting spinful fermions, our results demonstrate that when a repulsive Hubbard contact interaction between both species is switched on, two different delocalisation mechanisms emerge. One is the presence of the pair hopping and spin-flip terms, which lead to transport of the type identified in Refs.~\cite{Yu+18,Protopopov+19}, while the other arises due to the slow decay of the coupling constants in the transformed Hamiltonian at weak disorder, resulting in a large $l$-bit localisation length. The former mechanism is activated by the presence of $\mathrm{SU}(2)$ symmetry (either spin or pseudospin), while the latter is unrelated to symmetry and results from the interplay of the microscopic interactions $U/t$ and disorder strength $d/t$. 
While the existence of weak resonant terms in the presence of either $\mathrm{SU}(2)$ symmetry implies thermalisation in the long-time limit, they do not rule out the emergence of long-lived prethermal or glassy states  at large $d$ where non-ergodic features may persist for long times.
Further investigating the form of this relaxation, and particularly whether it exhibits glassy effects such as aging, may be an interesting topic for future work.

\emph{Conclusion} - In this work, we have investigated the role of continuous non-Abelian symmetries on the formation of local integrals of motion in the one-dimensional Fermi-Hubbard model, and shown the origin of a form of disorder-induced spin-charge separation. While these numerical results specify to one dimension, the qualitative form of the Hamiltonian does not depend on the dimensionality, and it seems reasonable to expect that the presence of a continuous non-Abelian symmetry will generically forbid a multi-component system from being written solely in terms of mutually commuting local integrals of motion in any dimension. This will prevent full many-body localisation, although there may exist atypical states which the resonant terms cannot act upon. These results obtained from the $l$-bits complement prior studies which examined the effects of $\mathrm{SU}(2)$ symmetry on the entanglement structure~\cite{Protopopov+17,Yu+18} and more general symmetry considerations~\cite{Potter+16}. Any perturbations which break the both spin and pseudospin $\mathrm{SU}(2)$ symmetries, no matter how slight (e.g. uncontrolled impurities or stray magnetic fields) will act to stabilise localisation.
It would be very interesting to apply similar techniques in the limit of strongly interacting spinful fermions, e.g. to the disordered $t$-$J$ model~\cite{Spalek88,Fu+18,Lemut+18} or $t$-$0$ model recently studied in the context of MBL~\cite{Bahovadinov+21,Kurlov+21}, and examine the effects of disorder in a sector where the clean system already exhibits strong spin-charge separation. In this regime, one might expect to see even more dramatic differences between spin and charge $l$-bits.

\emph{Acknowledgements} - This project has received funding from the European Union’s Horizon 2020 research and innovation programme under the Marie Skłodowska-Curie grant agreement No. 101031489 (Ergodicity Breaking in Quantum Matter). SJT acknowledges support from the NVIDIA Academic Hardware Grant Program, and thanks B.~Braunecker for insightful comments regarding normal-ordering corrections in the Hubbard model, and J.~Eisert for useful discussions. All data and code will be made available at~\cite{PyFlow,data}.

\bibliography{refs}

\pagebreak
\clearpage
\onecolumngrid
\appendix
\section{Supplemental Material to ``Disorder-induced spin-charge separation in the 1-D Hubbard model"}

\subsection{Tensor Flow Equation Method}

The method is explained in detail in Ref.~\cite{Thomson+21} and references therein, but we will recapitulate the key aspects here for the interested reader, and highlight a few differences that must be taken into account when dealing with spinful fermions. 
The main idea behind the method is to use a series of infinitesimal unitary transforms to diagonalise the Hamiltonian~\cite{Brockett91,Chu94,Glazek+93,Glazek+94,Wegner94,Wegner06,Kehrein07}. We parameterise the series of transforms using a fictitious `flow time' which we denote $l$ (where $l=0$ represents the basis of the initial microscopic Hamiltonian), and the infinitesimal transform which advances the Hamiltonian from $l \to l+ \ud l$ is given by:
\begin{align}
    \mathcal{H}(l + \ud l) &= \textrm{e}^{\eta(l) \ud l} \mathcal{H}(l) \textrm{e}^{-\eta(l) \ud l} \\
    &= \mathcal{H}(l) + \ud l [\eta(l),\mathcal{H}(l)],
\end{align}
where $\eta(l)$ is a suitable generator for the transform such that it removes off-diagonal terms in the $l \to \infty$ limit. This process can also be expressed in terms of an equation of motion, or `flow equation', for the running Hamiltonian:
\begin{align}
    \frac{\ud \mathcal{H}(l)}{\ud l} &=  [\eta(l),\mathcal{H}(l)].
\end{align}
By integrating this equation from $l=0$ to $l \to \infty$, we can continuously diagonalise the Hamiltonian. 
In this work we choose $\eta(l) = [\mathcal{H}_0(l),V(l)]$ where $\mathcal{H}_0(l)$ is the diagonal Hamiltonian at flow time $l$, and $V(l) = \mathcal{H}(l) - \mathcal{H}_0(l)$ is the off-diagonal part of the Hamiltonian at flow time $l$. This is also known as the Wegner generator -- or sometimes the canonical generator -- and its properties have been widely explored elsewhere in the literature, e.g. Refs~\cite{Wegner94,Kehrein07}. The main limitation of this generator is that it cannot remove off-diagonal terms which couple degenerate sectors of the Hamiltonian, as it works in a renormalisation group-like manner based on energy scale separation. (For example, this is why in the main text, the spin-flip term does not vanish in the case of charge disorder, as the spin-up and spin-down sectors are degenerate and the coupling between them cannot be removed.) The key difference with respect to previous flow equation works is that the microscopic model considered here contains two fermionic species, corresponding to spin-up and spin-down respectively. The fundamental (anti)commutation relations we will need are the following:
\begin{align}
    \{ c^{\dagger}_{i,\sigma}, c_{j,\sigma} \} &= \delta_{ij}, \\
    [AB,CD] &= A \{B,C \} D - C \{D,A \} B + C A \{B,D \} - \{C,A \} B D,
\end{align}
where the second identity is valid for any fermionic operators $A,B, C$ and $D$. Note that this identity is equivalent to $[AB,CD] = A [B,C] D - C [D,A] B + C A [B,D] - [C,A] B D$ (more convenient for bosons), which can be checked explicitly~\cite{Kehrein07}.

\begin{figure}[h!]
    \centering
    \includegraphics[width=0.7\linewidth]{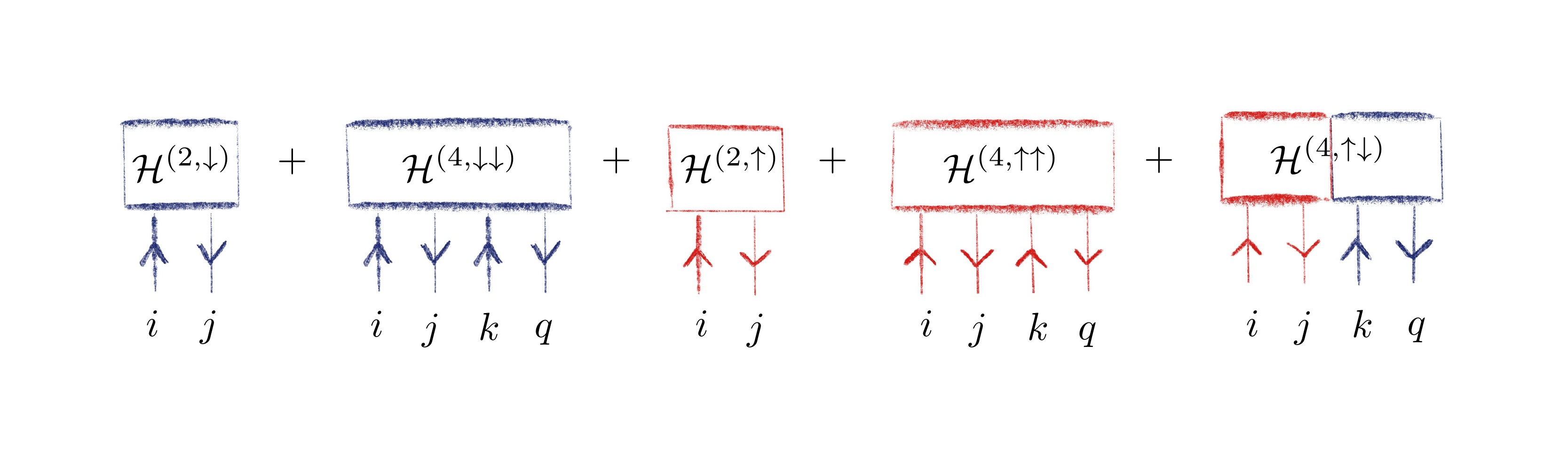}
    \caption{A graphical representation of a two-species Hamiltonian, with spin-up terms indicated in red and spin-down terms indicated in blue, showing the quadratic terms (two indices) as well as the interaction terms (four indices) contained in Eq.~\ref{eq.Hrun}.}
    \label{fig.H_graphical}
\end{figure}

Under the action of the Wegner flow, the running Hamiltonian is the following:
\begin{align}
    \mathcal{H}(l) = \sum_{ij} \sum_{\sigma = \uparrow, \downarrow} \mathcal{H}_{ij}^{(2,\sigma)}(l) :c^{\dagger}_{i,\sigma} c_{j,\sigma}: + \sum_{ijkq} \sum_{\sigma,\sigma'} \mathcal{H}_{ijkq}^{(4,\sigma,\sigma')}(l) :c^{\dagger}_{i,\sigma} c_{j,\sigma} c^{\dagger}_{k,\sigma'} c_{q,\sigma'}:,
    \label{eq.Hrun}
\end{align}
which is a summation over elements of a matrix plus elements of a rank-4 tensor. This form of continuous unitary transform generically creates all possible terms in the Hamiltonian which are allowed by symmetry, before sending the (non-degenerate) off-diagonal terms smoothly to zero, resulting in a diagonal final Hamiltonian (where `diagonal' means it can be written in terms of fermionic number operators), in the absence of degenerate sectors. For $l=0$, Eq.~\ref{eq.Hrun} reduces to the initial microscopic Hamiltonian. The running Hamiltonian can be represented using the graphical notation shown in Fig.~\ref{fig.H_graphical}
where the up and down arrows refer to creation and annihilation operators respectively, and for convenience the different spin species are indicated in different colours to make the structure of the model clearer. Note that in principle there could be an interaction term $\mathcal{H}^{(4,\downarrow \uparrow)}$ but we can always rearrange the operators and combine it with the $\mathcal{H}^{(4,\uparrow \downarrow)}$ term shown above in order to save the mixed-spin terms as a single array in the memory.

\begin{figure}[h!]
    \centering
    \includegraphics[width=0.8\linewidth]{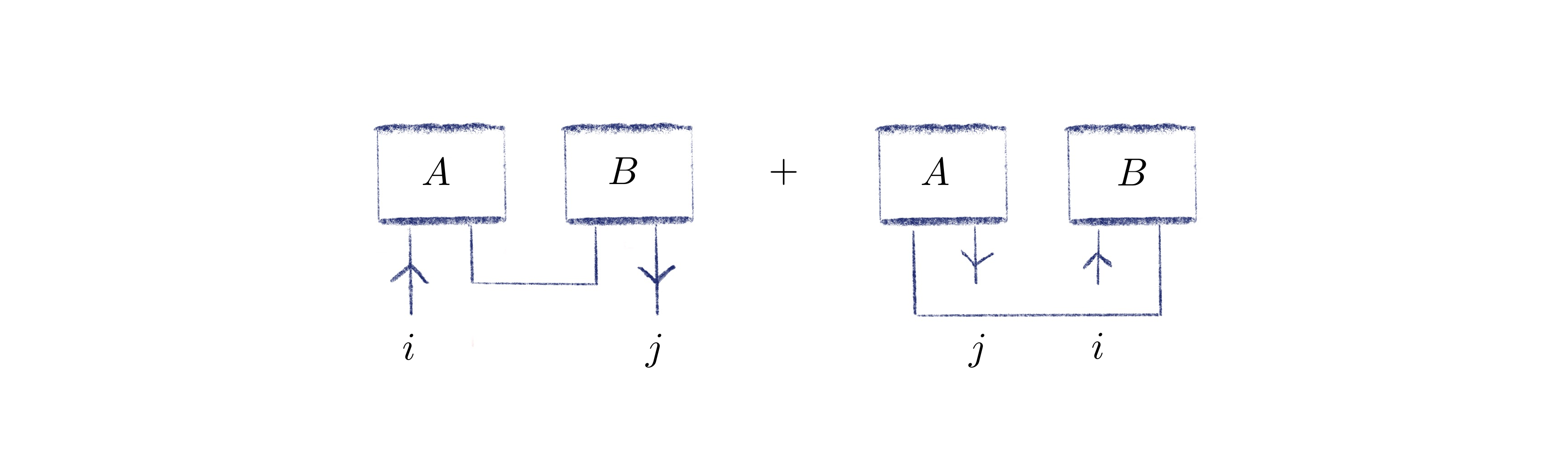}
    \caption{A graphical representation of the commutator $[A,B]$ for two matrices $A$ and $B$, where the summation over a shared index is represented as the contraction of two legs of the matrices.}
    \label{fig.matmul}
\end{figure}

\begin{figure}[h!]
    \centering
    \includegraphics[width=0.8\linewidth]{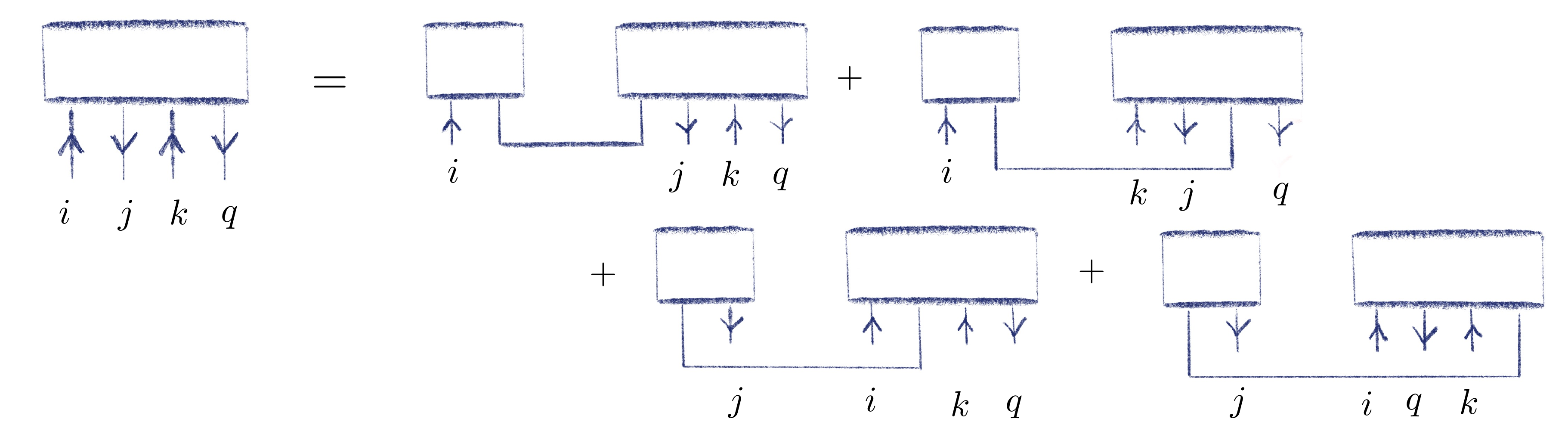}
    \caption{The one-point contractions which constitute the commutator of a square matrix and a rank-4 tensor, resulting in another rank-4 tensor. The indices on the right hand side must be correctly permuted using fermionic anticommutation relations in order to add all the contributions into a single rank-4 tensor.}
    \label{fig.matmul42}
\end{figure}

Every stage of the flow equation procedure can be recast in this way, with the generator acquiring a similar graphical structure, and also any transformed operator can be written in a similar way (although note that creation/annihilation operators will contain vectors/tensors with an \emph{odd} number of indices, while number operators will contain matrices/tensors with an \emph{even} number of indices), and so this structure allows us to perform all of the computations required to compute and apply infinitesimal unitary transforms. All of the commutators we are required to calculate can be recast in this notation as matrix/tensor contractions. For example, the commutator of two (fermionic) matrices $A$ and $B$ can be written as a sum of all possible one-point contractions:
\begin{align}
    [A,B] = \sum_{ijk} (A_{ik}B_{kj} + A_{kj} B_{ik}),
\end{align}
(note that upon permuting the indices using fermionic anticommutation relations this becomes the more conventional $\sum_{ijk} (A_{ik}B_{kj} - B_{ik}A_{kj})$) which can be represented graphically as shown in Fig.~\ref{fig.matmul}.
The higher-order terms can also be represented similarly as a sum of one-point contractions, as shown in Fig.~\ref{fig.matmul42},
where the indices can be permuted back into the order $(i,j,k,q)$ using standard anticommutation relations, recalling that swapping the order of two normal-ordered operators does not generate any new lower-order terms~\cite{Kehrein07}. At this point, however, the curious reader may note that it's also possible to perform \emph{two-point} contractions without reducing everything to an unimportant constant, and may be wondering what these contractions would look like and in what cases they are necessary. It turns out that these two-point contractions constitute the normal-ordering corrections which are used in the main text and which we shall discuss in the next section.


\subsection{Normal-Ordering Corrections}

A key ingredient of the method is normal-ordering, which allows corrections from high-order terms to be resummed and included in lower-order terms. It is precisely this feedback from high- to low-order terms that enables this method to be a powerful non-perturbative tool, capable in some cases of reproducing results from one-loop and two-loop renormalisation group calculations. For full details, see Ref.~\cite{Kehrein07}. Here, we will simply recap the most relevant details for the present work. For example, the commutator of a 4-body normal-ordered fermionic string with a 2-body normal-ordered fermionic string is given by:
\begin{align}
[:c^{\dagger}_{\alpha} c_{\beta} c^{\dagger}_{\gamma} c_{\delta}: , :c^{\dagger}_{\mu} c_{\nu}:] = &\phantom{+} -(G_{\alpha \nu}+\tilde{G}_{\nu \alpha}) :c^{\dagger}_{\mu} c_{\beta} c^{\dagger}_{\gamma} c_{\delta}: - (G_{\gamma \nu}+\tilde{G}_{\nu \gamma})  :c^{\dagger}_{\alpha} c_{\beta} c^{\dagger}_{\mu} c_{\delta}: \nonumber \\
& + (\tilde{G}_{\beta \mu} + G_{\mu \beta}) :c^{\dagger}_{\alpha} c_{\nu} c^{\dagger}_{\gamma} c_{\delta}: + (\tilde{G}_{\delta \mu} + G_{\mu \delta}) :c^{\dagger}_{\alpha} c_{\beta} c^{\dagger}_{\gamma} c_{\nu}: \nonumber \\
&+ (G_{\alpha \nu} \tilde{G}_{\beta \mu} - G_{\mu \beta} \tilde{G}_{\nu \alpha} ):c^{\dagger}_{\gamma} c_{\delta}:+ (G_{\alpha \nu} \tilde{G}_{\delta \mu} - G_{\mu \delta} \tilde{G}_{\nu \alpha} ):c^{\dagger}_{\gamma} c_{\beta}:  \nonumber \\
& + (G_{\gamma \nu} \tilde{G}_{\beta \mu} -G_{\mu \beta} \tilde{G}_{\nu \gamma}) :c^{\dagger}_{\alpha} c_{\delta}: +  (G_{\gamma \nu} \tilde{G}_{\delta \mu} -G_{\mu \delta} \tilde{G}_{\nu \gamma}) :c^{\dagger}_{\alpha} c_{\beta}: ,
\end{align}
where $G_{\alpha \beta} = \langle c^{\dagger}_{\alpha} c_{\beta}\rangle$, and $\tilde{G}_{\beta \alpha} =\langle c_{\beta} c^{\dagger}_{\alpha}\rangle$. The first two lines are the standard result which can be represented as the one-point contractions shown in the previous section (as $G_{\alpha \beta} + \tilde{G}_{\beta \alpha} = \delta_{\alpha \beta}$), and the final two lines are the normal-ordering corrections which depend explicitly on the chosen reference state. Note that in this case, they do not simply give delta functions and allow us to directly sum over the indices, but instead give a result which also contains an expectation value which must be computed with respect to some reference state. For a product state, this gives:
\begin{align}
    G_{\alpha \beta} \tilde{G}_{\delta \gamma} - G_{\gamma \delta} \tilde{G}_{\beta \alpha} = \delta_{\alpha \beta} \delta_{\gamma \delta} (\langle n_{\alpha} \rangle - \langle n_{\gamma} \rangle),
    \label{eq.norm}
\end{align}
which vanishes for a homogeneous reference state, but can give a non-zero contribution otherwise. The normal-ordering corrections can be represented graphically as a sum of all possible two-point contractions, where it must be understood that this is a convenient notation but that in any computational implementation this is not simply a matrix/tensor contraction and must also include multiplication of the elements by the expectation values above.
\begin{figure}[h!]
    \centering
    \includegraphics[width=0.8\linewidth]{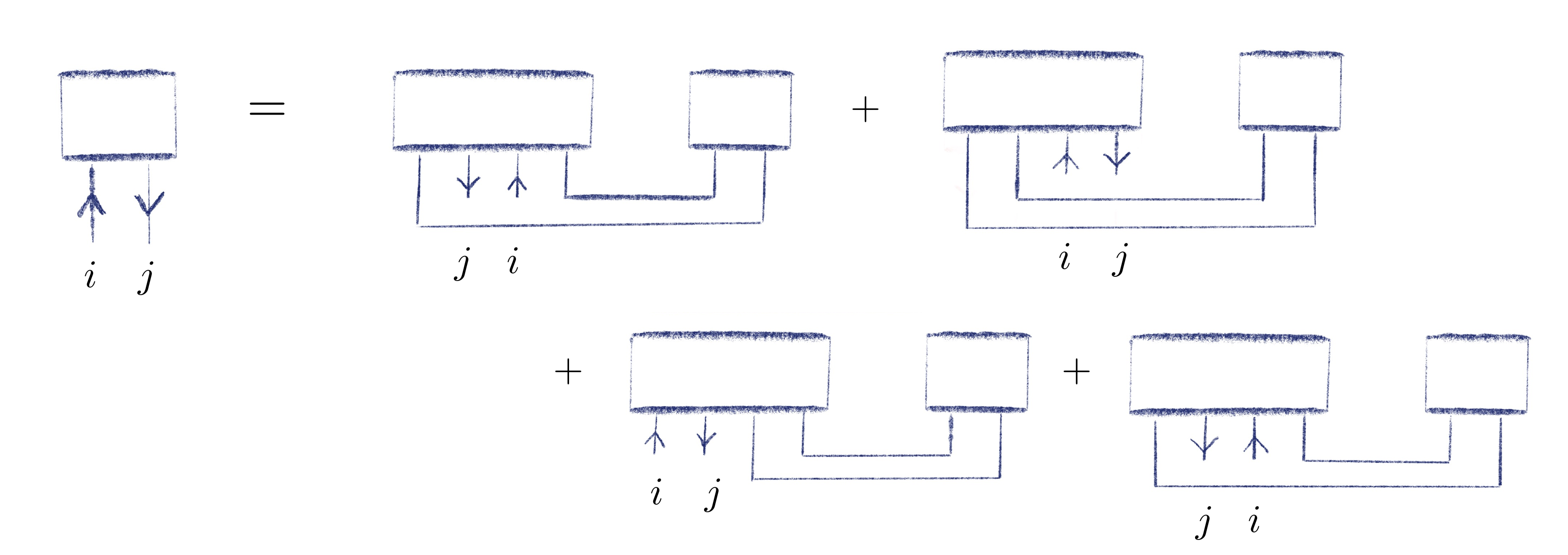}
    \caption{The two-point contractions involved in computing the normal-ordered commutator of a square matrix with a rank-4 tensor, resulting in a square matrix with two free indices.}
    \label{fig.NO}
\end{figure}

This procedure provides an efficient, easy-to-visualise way of computing sophisticated normal-ordering corrections of many-body quantum systems. A similar principle holds for higher-order terms, which we shall not show here for brevity, as using the graphical notation they can be straightforwardly worked out. For normal-ordering corrections associated to the product of two fourth-order terms, note that this involves $\mathcal{O}(L^6)$ floating point operations per flow time step. The method incurs a total memory cost of $\mathcal{O}(q \times L^4)$ where $q \sim 10^3$ is the total number of flow time steps. (Although this is only necessary if one wishes to store the unitary transform, e.g. for computing the $l$-bits or performing time evolution.) For completeness, we note that for terms sufficiently many fermionic operators, 3-point and higher contractions are also possible, but we do not consider them in the present work.

In the present case, normal-ordering is important for two main reasons. Firstly, it provides a consistent way of ordering operators which is necessary for the procedure to remain well-defined and important for the computational implementation~\cite{Thomson+21}. Secondly, it allows for the generation of same-species interaction terms which would not appear in a purely perturbative treatment using continuous unitary transforms (e.g. the terms $\tilde{\Delta}_{ij}^{\uparrow / \downarrow}$ would be zero, resulting in $\Omega_{ij} =0$ and $|\Delta_{ij}^{\rho}| = |\Delta_{ij}^{\sigma}|$). The mixed-species anticommutation relations can be taken into account straightforwardly by graphically computing the relevant commutators starting from Fig.~\ref{fig.H_graphical} and applying the logic demonstrated in Figs.~\ref{fig.matmul}, \ref{fig.matmul42} and \ref{fig.NO}, recalling that only indices of the same colour may be contracted with one another. The complex problem of computing high-order normal-ordered commutators becomes a simpler combinatoric exercise of connecting all possible pairs of legs, which is far less algebraically tedious and easier to spot errors. In the future it may even be possible to automate this process, further improving the computational implementation.

The normal-ordering corrections with respect to the mixed interaction vertex are particularly important, as these give rise to the same-species interaction terms which lead to the spin-charge separation shown in the main text. For example, the spin-up and spin-down interactions terms are initially generated by the following normal-ordering corrections, before acquiring their own dynamics under the action of the Wegner generator:

\begin{figure}[h!]
    \centering
    \includegraphics[width=0.6\linewidth]{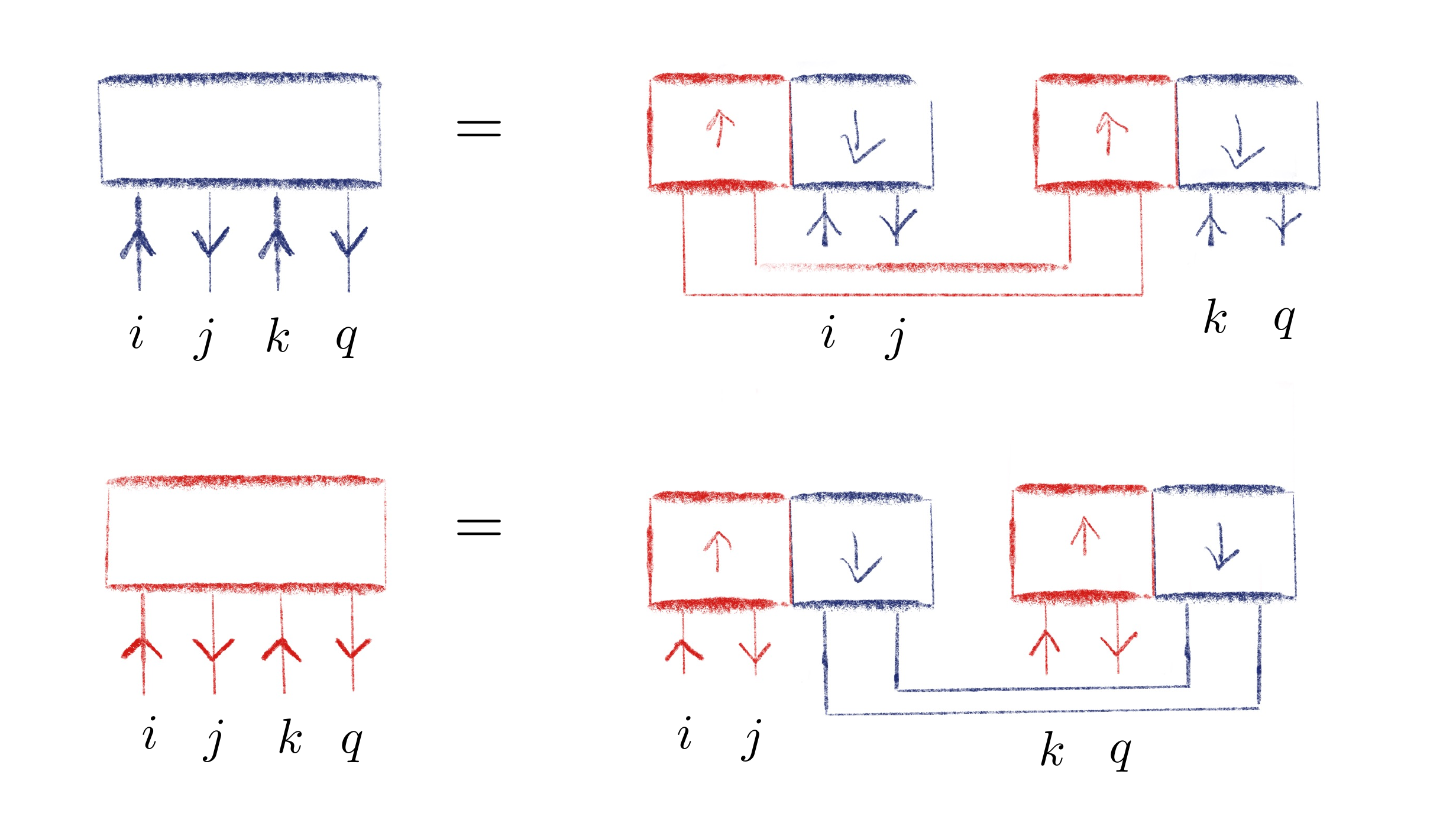}
    \caption{A sketch of the normal-ordering corrections resulting from two-point contractions of the interaction terms in the Hamiltonian, which give rise to the same-species interactions. As detailed in the text, these corrections depend explicitly on the chosen reference state, and are zero for the vacuum.}
    \label{fig.NO44}
\end{figure}

It bears emphasising that these corrections depend crucially on the choice of reference state. For example, if one chooses the vacuum (the conventional choice in quantum field theory calculations~\cite{Altland+10}) then the expectation values in Eq.~\ref{eq.norm} will be zero and this same-species interaction will not be generated during the flow.

It is also worth mentioning that any symmetry breaking in the reference state will be imposed upon the final Hamiltonian. In the main text, for both types of disorder we choose a state which is a charge density wave in the diagonal basis (see next section for information on how this is computed), which is a generic choice for a highly excited state. This state does, however, break the $\mathrm{SU}(2)$ pseudospin symmetry, however as the corrections resulting from this state are quantitatively small, this symmetry breaking does not have significant physical consequences, except that the case of spin disorder the coupling constants $U^{*}_{ij}$ are not strictly zero to within machine precision, but instead reach a maximum value of around $\sim 10^{-10}$ at a disorder strength $d/t=2.0$. In the case of a reference state which is a spin density wave in the diagonal basis, $\ket{\uparrow \downarrow \uparrow \downarrow...}$ which explicitly breaks the $\mathrm{SU}(2)$ symmetry, we find qualitatively similar behaviour as shown in the main text (except that now in the case of charge disorder, the spin-flip coefficient $\overline{U}_{ij}$ is not strictly zero, but again takes a maximum value of order $\sim 10^{-10}$).

\begin{figure}
    \centering
    \includegraphics[width=\linewidth]{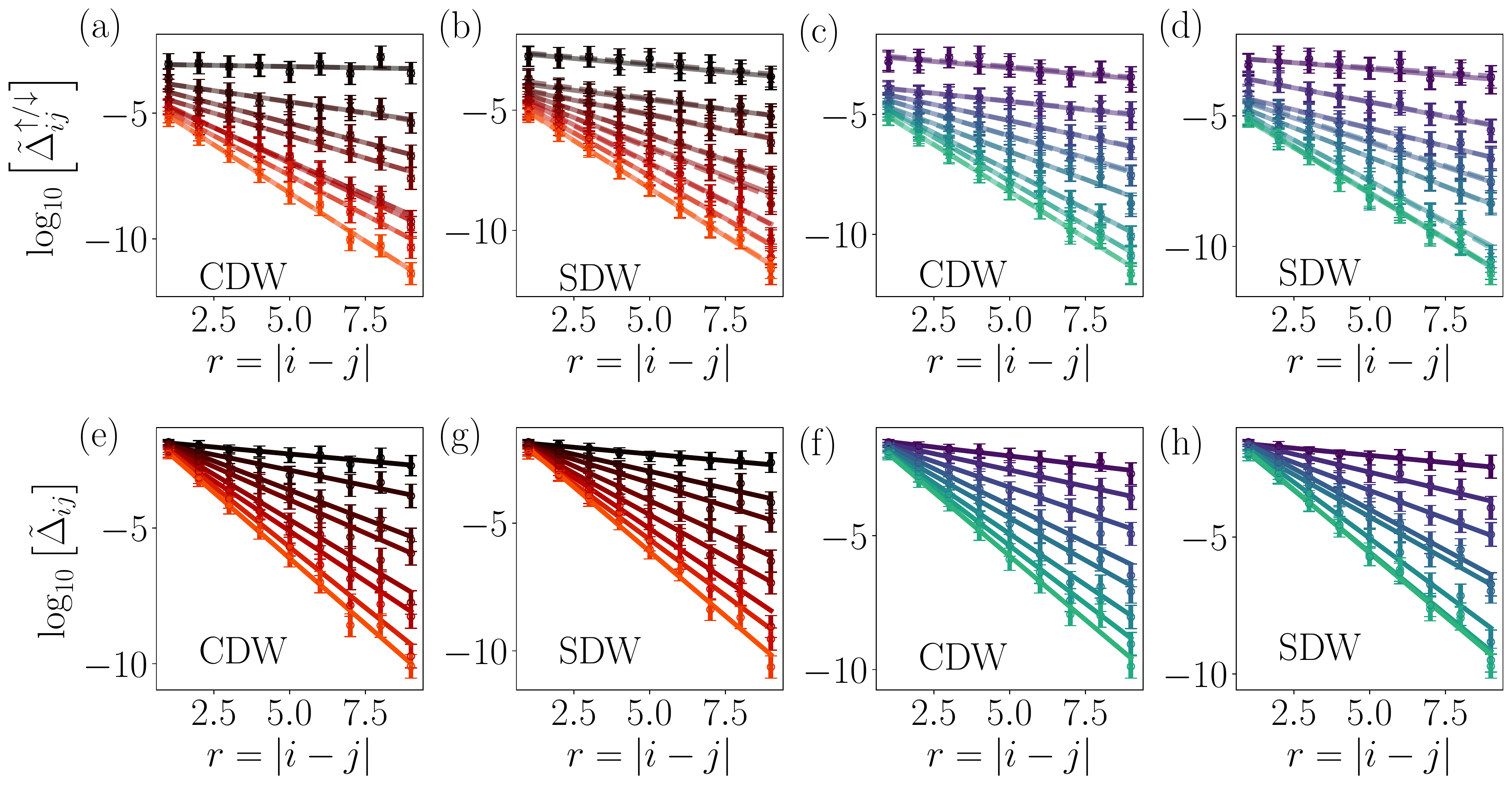}
    \caption{Comparison demonstrating the effects of different choices of normal ordering states, for $L=10$ and $N_s=60$. The two states used are the charge density wave (CDW) and spin density wave (SDW) states. The colour coding follows the convention of the main text, indicating charge disorder in red/black and spin disorder in green/blue. The quantities shown here are the same as computed in Fig. 2 of the main text.}
    \label{fig.no_state}
\end{figure}

A comparison of the quantities shown in Fig. 2 of the main text for different choices of normal-ordering state is shown in Fig.~\ref{fig.no_state}, where we see that the choice of normal-ordering state does not result in significant differences in the resulting coupling coefficients. The $\tilde{\Delta}^{\uparrow/\downarrow}_{ij}$ terms show the greatest variation, as they are generated solely by these correction terms, however the differences between them are within the error bars.

\subsection{Scale-Dependent Normal-Ordering}

One subtlety usually ignored when computing normal ordering corrections is that strictly speaking, the corrections should be evaluated in the instantaneous basis, i.e. they explicitly depend on the value of the flow time $l$. Being able to implement this requires complete knowledge of how the reference state evolves under the action of the transform, which in general is not known, and as corrections due to the scale-dependence can be shown to be subleading~\cite{Kehrein07}, they can usually be safely neglected. 

In the current approach, however, it is possible to implement this \emph{exactly} (even for excited states) as long as our reference state is any state of the non-interacting system. As the system sizes we study are relatively small, and diagonalising the non-interacting Hamiltonian is not a computationally costly step, we can at each flow time step fully diagonalize the free Hamiltonian $\mathcal{H}^{(2,\sigma)}$ for $\sigma = \uparrow,\downarrow$, allowing us to exactly obtain the $l$-dependent eigenstates. As the free Hamiltonian is Anderson-localised and each of the eigenstates are exponentially localised around each of the real-space lattice sites, constructing the state which reads $\ket{\tilde{\psi}} = \ket{0101...}$ (for example) in the diagonal basis simply requires extracting the single-particle eigenstates which are peaked on lattice sites on which there are $1$s, and adding them together with appropriate normalisation. We illustrate this with the following pseudo-code for spinless fermions, using a notation close to Python.

\begin{algorithm}
    \SetKwInOut{Input}{Input}
    \SetKwInOut{Output}{Output}

    \underline{function normstate} $(H)$\;
    \Input{Non-interacting Hamiltonian $H(l)$ as an $L \times L$ matrix}
    \Output{Density expectation values in reference state at scale $l$, equal to a normalised CDW in the $l \to \infty$ basis}
    import numpy as np\;
    \_,V = np.linalg.eigh(H)\;
    state = np.zeros(n)\;
    sites = range(0,n,2)\;
    \For{site in sites}{
        \For{i in range(n)}{
            \If{np.argmax(np.abs(V1[:,i]))==site}{
                psi = V[:,i]\\
                state += np.array([v**2 for v in psi])\\}
        }}
    state *= 1/(n/2)\\
    return state
    \caption{Scale-dependent normal ordering w.r.t. an excited state of the free Hamiltonian}
\end{algorithm}

This technique allows us to compute normal-ordering corrections with respect to arbitrary excited states, and turn a challenging limitation of conventional flow equation methods into an straightforwardly computable numerical tool which allows for the incorporation of highly non-trivial, non-perturbative effects in many-body quantum systems.

\subsection{Basis Transformation}

We can transform any operator initially defined in the diagonal basis into the original microscopic basis using the flow equation:
\begin{align}
    \frac{\ud \tilde{O}(l)}{\ud l} = [\eta(l),\tilde{O}(l)],
\end{align}
where $\eta(l)$ is the Wegner generator computed and stored during the initial diagonalisation process. In practice, this corresponds to `reversing the flow' and applying the inverse unitary transform originally used to diagonalise the Hamiltonian. Storing the generator at all flow time steps $l$ allows the evolution of any operator to and from the diagonal basis, using a similar matrix/tensor notation as we use to represent the Hamiltonian. For example, the $l$-bit number operator will evolve to have the same structure as the Hamiltonian in Fig.~\ref{fig.H_graphical}, with spin-up, spin-down and mixed-spin blocks which all enter into the commutation relations used to compute the flow of the operator.
Note that while the $l$-bits should by design commute with the Hamiltonian, two comments are in order. Firstly, they will only commute with the Hamiltonian up to corrections on the order of the neglected terms in the running Hamiltonian, therefore it is important that these neglected terms are small (as they are here in the case of weak interactions). Secondly, although rarely an issue, one should keep in mind that strictly speaking, employing a normal-ordering procedure to truncate the operator expansion imposes only the more relaxed constraint that they obey canonical commutation relations computed with respect to the reference state chosen for the normal-ordering - see Ref.~\cite{Kehrein07} for further details. 

\subsection{Accuracy Benchmarks: Comparison with Exact Diagonalisation}

\begin{figure}
    \centering
    \includegraphics[width=0.65\linewidth]{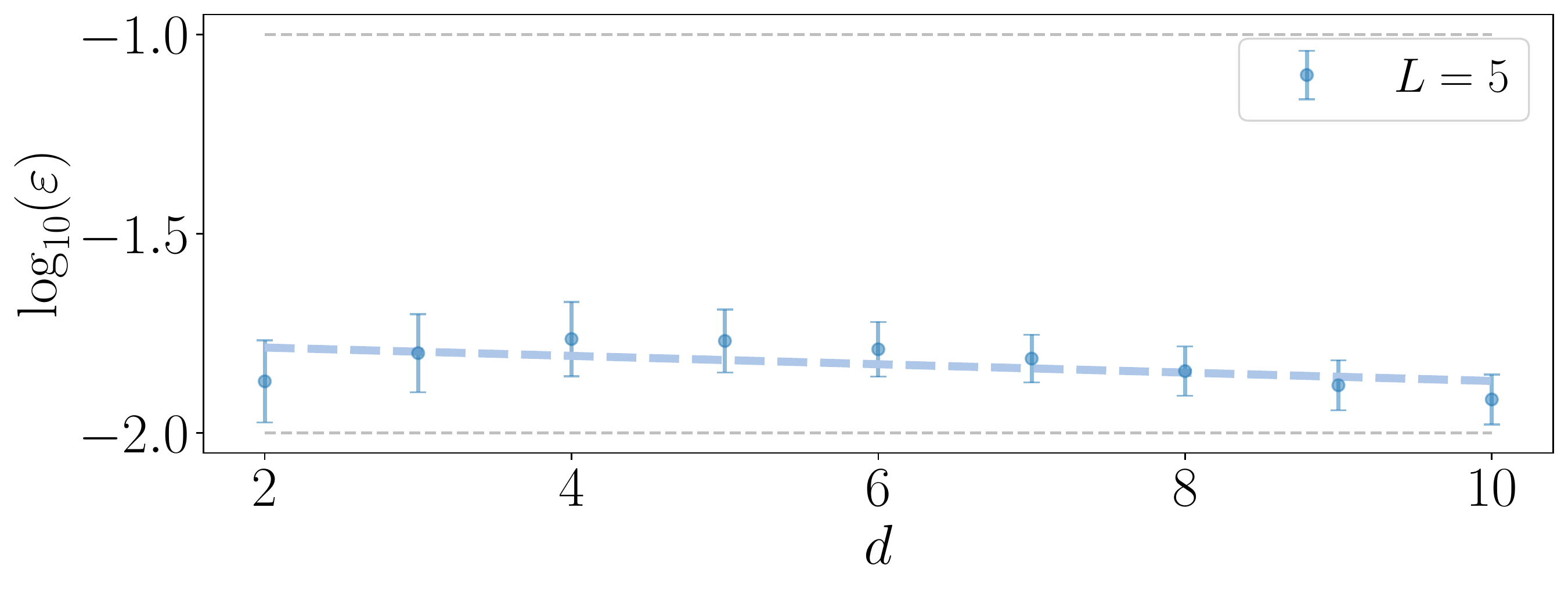}
    \caption{Median relative error in the eigenvalues obtained using the flow equation method, compared with exact diagonalisation for two system size $L=5$ and using $N_s = 1920$. The error bars represent the median average deviation. Dashed lines are guide to the eye at relative errors of $10\%$ and $1\%$ respectively. The highest error shown is $1.72\%$, at disorder strength $d/t=4.0$.}
    \label{fig.rel_err}
\end{figure}

Although the focus of the main text is the qualitative form of the fixed-point Hamiltonian obtained using the TFE method, in order to confirm the reliability of the method we can examine some brief benchmarks. As a first -- rather crude -- measure, we can compare the accuracy of the eigenvalues obtained from the fixed-point Hamiltonian with those obtained from exact diagonalisation (ED), using the \texttt{QuSpin} package~\cite{Weinberg+17,Weinberg+19}. Because of the computational cost of exactly diagonalising a system of spinful fermions, here we focus only on very small system sizes. In order to have a meaningful benchmark, we specify to the case of spin disorder where the TFE method can fully diagonalise the Hamiltonian.

Here we compute the relative error in the eigenvalues obtained from the TFE method and from ED. The eigenvalues can be obtained in the TFE method by applying the diagonal fixed-point Hamiltonian to all of the product states in the basis. The relative error in the $n^{\textrm{th}}$ eigenvalue (for a given disorder realisation) is defined as:
\begin{align}
    \epsilon_n =  \left| \frac{\epsilon_n^{FE} - \epsilon_n^{ED}}{\epsilon_n^{ED}} \right|,
    \label{eq.rel_err}
\end{align}
and we define $\varepsilon$ as the median error across all eigenstates and disorder distributions. We use the median rather than the mean, as strictly speaking our normal-ordering procedure means that we target an effective Hamiltonian valid for highly excited states only, and as such the effective Hamiltonian we obtain should not be expected to accurately reproduce all eigenvalues. In particular, the low-energy eigenvalues are not likely to be captured accurately. In addition, many of the eigenvalues are close to zero, resulting in a small number of large outliers which contaminate the mean error (e.g. if an eigenvalue is close to zero, then the factor of $\epsilon_n^{ED}$ in the denominator of can lead to a small number of unrepresentatively large relative errors), but have a less significant effect on the median.

The results are shown in Fig.~\ref{fig.rel_err} for system size $L=5$ (with Hilbert space dimension $1024$) at fixed interaction strength $U=0.1$, averaged over $N_s=1920$ disorder realisations. The median relative error is on the order of one percent and roughly constant -- with perhaps a slow reduction as the disorder strength is increased -- confirming that the method is reliable.

\subsection{Accuracy Benchmarks: Invariants of the Flow}

\begin{figure}
    \centering
    \includegraphics[width=0.7\linewidth]{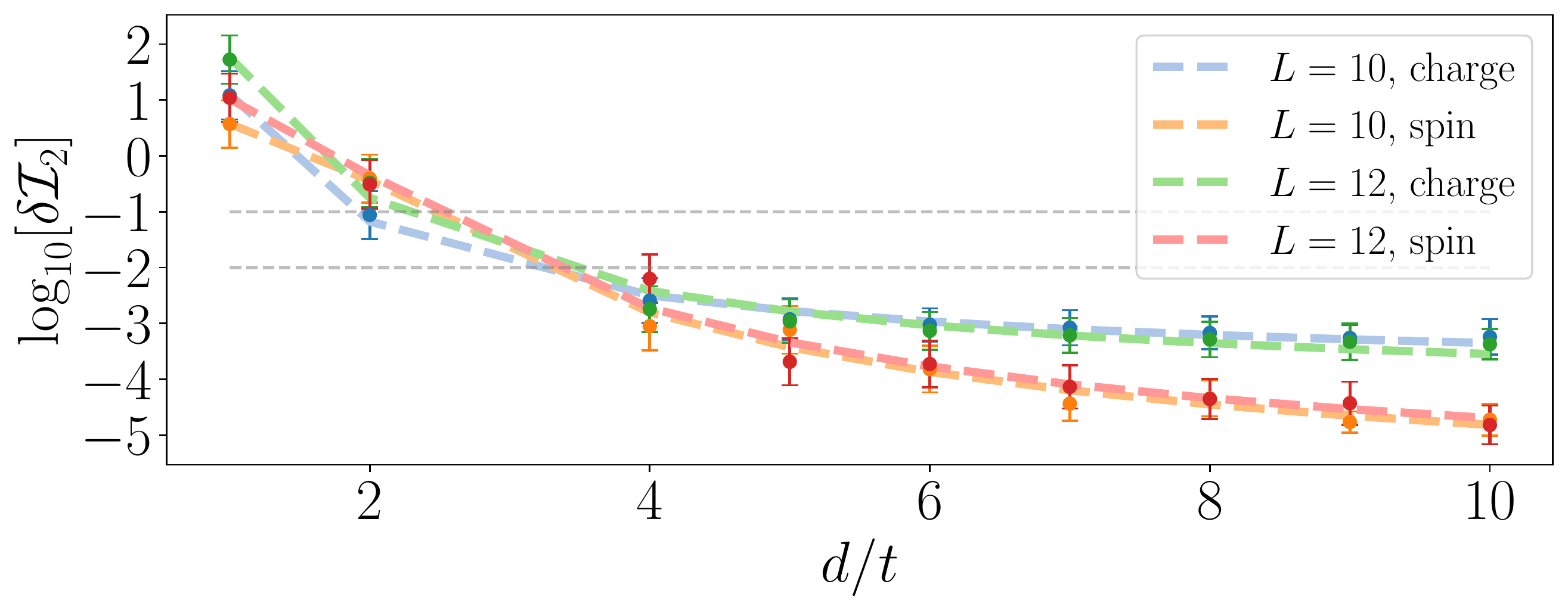}
    \caption{Conservation of the flow invariant, demonstrating that the transform remains unitary at strong disorder but deviates for very weak disorder, as expected. Results are shown for both charge and spin disorder. Dashed grey lines are guides to the eye at $\delta \mathcal{I}_2 = 10^{-1}$ and $\delta \mathcal{I}_2 = 10^{-2}$, representing deviations from unitarity on the order of $10\%$ and $1\%$ respectively.}
    \label{fig.SM_inv}
\end{figure}

The above comparison with ED is useful, but clearly limited - as our normal-ordering procedure specifies to a particular class of excited states, it cannot be expected to give the entire energy spectrum accurately, cannot be used in the case of charge disorder (where the Hamiltonian is not fully diagonalised, and the remnant off-diagonal terms cannot be straightforwardly taken into account without building the Hamiltonian as a full matrix), and moreover this comparison is only possible for small enough system sizes to be amenable to full exact diagonalisation.

A second, independent benchmark of the method is to look at so-called `invariants of the flow'. The TFE method is based around a unitary transform, which has to conserve certain properties of the Hamiltonian. In particular, it should conserve traces of integer powers of the Hamiltonian:
\begin{align}
    \mathcal{I}_p(l) = \textrm{Tr} \left[ \mathcal{H}^p \right].
\end{align}
By comparing this transform at the start ($l=0$) and end ($l \to \infty$) of the flow, we can get a self-consistent measure of how close to unitary the overall transform is, giving us a way to measure errors due to the truncation of the running Hamiltonian or any divergences that may occur in the delocalised phase. 

In certain cases, it is possible to obtain analytical expressions for the invariants of the flow~\cite{Monthus16}, however here we will proceed to compute the flow invariant numerically. We will specify to the case of $p=2$, and we will perform the trace over all states with two spin-up fermions and two spin-down fermions. We could in principle use any orthonormal basis, but we need to choose a small enough basis that it is feasible to construct the Hamiltonian numerically - this is the smallest non-trivial basis that permits us to take into account both same-species interactions and inter-species interactions. We take the Hamiltonian used in the TFE method and build it as a matrix in the chosen basis using \texttt{QuSpin}~\cite{Weinberg+17,Weinberg+19}, then compute the trace of $\mathcal{H}^2$ numerically.

The quantity of interest is the normalised difference between the second invariant at the start and end of the flow:
\begin{align}
    \delta \mathcal{I}_2 = \left| \frac{\mathcal{I}_2(l=0) - \mathcal{I}_2(l \to \infty)}{\mathcal{I}_2(l=0)} \right|.
\end{align}
If $\delta \mathcal{I}_2$ is small, then the flow is (close to) unitary, whereas if it is large then the transform has significantly deviated from unitary. We would expect the invariants of the flow to be well-conserved for weak interactions and/or strong disorder, but to diverge in the delocalised phase where the ansatz for the form of the running Hamiltonian is no longer sufficient to capture the delocalised nature of the excitations. This is precisely what we see in Fig.~\ref{fig.SM_inv}, which shows the median flow invariant for the dataset considered in the main text, with $L=12$, $U/t=0.1$ and $N_s=128$, for both charge and spin disorder. At low disorder strengths, the invariant diverges -- signifying a likely delocalisation transition -- but at strong disorder the invariant is conserved to an accuracy of approximately one percent or better. Results are also shown for a smaller system of size $L=10$ with $N_s=60$, demonstrating qualitatively similar behaviour. (Note that the divergence found here at the lowest disorder strength is also visible in Fig.~4c) of the main text: for $d/t=1.0$, the interaction terms in the $l$-bit are clearly orders of magnitude larger than for all other disorder strengths.)

\end{document}